# Vortex Interaction in Triple Flickering Buoyant Diffusion Flames

Tao Yang [a], Yicheng Chi [a], Peng Zhang [a,*]

[a] *Department of Mechanical Engineering, The Hong Kong Polytechnic University, Hung Hom, Kowloon, Hong Kong*

**Abstract**
Triple flickering buoyant diffusion flames as a nonlinear dynamical system of coupled oscillators were computationally investigated. The four distinct dynamical modes (in-phase, death, rotation, and partial in-phase) observed in the previous candle-flame experiments were computationally reproduced for jet diffusion flames of methane. The four modes were interpreted from the perspective of vortex interaction and particularly of vorticity reconnection and vortex-induced flow. Specifically, the in-phase mode is caused by the periodic shedding of the trefoil vortex formed by the reconnection of three toroidal vortices; the death mode is due to the suppression of vortex shedding at small Reynolds numbers; the rotation mode appears as three toroidal vortices alternatively shed off with a constant phase difference; the partial in-phase model is caused by the vorticity reconnection of two toroidal vortices leaving another one shedding off in anti-phase.

## 1. Introduction

Nonlinear dynamics of coupled oscillators is a long-lasting problem in the science of complex system[1, 2]. In recent years, flickering buoyant diffusion flames treated as a type of nonlinear oscillators have gained increasing attentions[3-14]. Flicker of a candle flame or a wood fire is a familiar periodic phenomenon, in which a portion of luminous flame stretches upwardly and pinches off from the flame, leading to a rising flame "bubble" that quickly burns out. In the First Combustion Symposium, Chamberlin and Rose[15] reported the flicker of Bunsen diffusion flames and found that the flicker frequency of ten times per second was not greatly affected by the flame conditions. Similar phenomenon was also discovered in a Burke-Schumann butane diffusion flame [16] and in pool fires [17, 18]. where it was named as "flame puffing".

Many studies have attempted to understand the physics of flickering diffusion flames. They pointed to a striking feature of the flames that flicker is not caused by an externally forced vibration or by the alternate flame extinction and re-ignition. In fact, the flicker of diffusion flames is a self-exciting flow oscillation. A prominent substantiation of the feature was owing to Chen et al.'s[19] flow visualization of a methane jet diffusion flame, in which the small-scale vortices inside of the luminous flame are due to the Kelvin-Helmholtz instability of the fuel jet, and the large-size toroidal vortices (a.k.a. vortex rings) outside the luminous flame are due to the buoyance-induced Kelvin-Helmholtz instability. Buckmaster and Peters[20] conducted a linear stability analysis to a self-similar solution of an annular burner diffusion flame and established a correlation between the Kelvin-Helmholtz instability of the buoyancy-induced flow and the flicker frequency. Their theoretical predictions were improved by Moreno-Boza et al.[6] by using a global linear stability analysis.

In order to overcome the limitation of linear stability analysis in describing the nonlinear phenomenon of flickering flames, Xia and Zhang [11] focused on the formation, growth, and shedding of a toroidal vortex within a flicker period. They obtained an analytical formula, $\Gamma(\tau) = C_h Ri\, St^{-2} + C_j Fr^{-1/2} St^{-1}$, where $\Gamma(\tau)$ is the dimensionless circulation of the toroidal vortex at the end of the period $\tau$, $St = fD/V$ is the Strouhal number, $Fr = V^2/gD$ is the Froude number, $Ri = (\rho_\infty/\rho - 1)gD/V^2$ is the Richardson number; $f$ is the flicker frequency, $g$ is the gravitational constant, $D$ the fuel inlet diameter; $\rho$ and $\rho_\infty$ are the density of flame and ambient, respectively; $C_h$ and $C_j$ are the experimentally determined correction constants for the advective speed of the vortex and the circulation addition by the inflow, respectively. By applying a vortex shedding criterion, $\Gamma(\tau) = C$, where $C$ is a system-dependent constant[21-25], they obtained an analytical solution of $St$ that generalizes the previous scaling laws[17-23] and well predicts experimental data in the literature for $Fr \ll 1$ and $Ri \gg 1$.

Spontaneous synchronization is an interesting dynamical behavior of coupled oscillators. Kitahata et al.[26] observed that two flickering candle flames exhibit transition from in-phase synchronization to anti-phase synchronization by increasing the distance between the candles. Similar phenomena were also reported by Forrester[27] and Manoj et al.[28] for candle flames. Dange et al.'s[29] high-speed shadowgraph and CH* chemiluminescence indicated that the interaction between buoyance-induced vortices might play a significant role in producing the distinct dynamical modes. This vortex-dynamical viewpoint was also supported by Fujisawa et al.'s PIV velocity field measurement of pipe-burner diffusion flames[30] and by Bunkwang et al.'s experiments of methane/air jet diffusion flames[31, 32].

The in-phase and anti-phase modes of dual buoyant flickering diffusion flames was numerically reproduced by Yang et al.[33] for small-scale heptane pool flames. Their results show that the interaction of two flickering flames at relatively larger separation distance is through the flow induced by each other's vortex while the interaction at relatively smaller separation distance is through the viscous diffusion and cancellation of vorticity in the region between two flames. In their vortex-dynamical interpretation, the transition from in-phase mode to anti-phase mode is controlled by a single dimensionless parameter, $\alpha Gr^{1/2}$, where $\alpha = L/D$ is the dimensionless flame separation distance, and $Gr = gD^3/\nu$ is the Grashof number. This one-parameter criterion for mode transition unifies their computational results for differential pool separation $L$, pool diameter $D$, gravitational constant $g$, and ambient viscosity $\nu$.

Recently, the collective behaviors of coupled multiple flickering diffusion flames have been experimentally studied by using candle flames in various geometrical arrangements[26-29, 34-38]. Okamoto et al.[34] investigated three flames in an equilateral triangle arrangement. Manoj et al.[38] experimentally observed very rich dynamical modes in various flame arrangements such as straight line, triangle, square, star, and annular networks. Forrester[27] observed that a ring of flames collectively enhance or suppress the height of a central flame. The most fascinating discovery in these studies is the spontaneous symmetry breaking due to nonlinear coupling of flame oscillators. For example, in an equilateral triangle arrangement with $D_3$ symmetry, besides the symmetrical in-phase mode (three flames flicker without phase difference), two asymmetrical modes were identified such as the partial in-phase mode (two in-phase flames are anti-phase with the third one) and the rotation mode (three flames flicker with a phase difference of $2\pi/3$). In addition, the death mode was observed as three flames cease to oscillate. Okamoto et al.[34] found the symmetric Hopf bifurcation theory can partially explain the existence of four distinct dynamical modes. They also proposed a vortex-dynamical conjecture on the physical mechanism of the triple flickering flame system, which however is quite hypothetical and lacks details.

Inspired by the previous study of Yang et al.[33], the present work aimed to interpret the four dynamical modes (i.e. in-phase, partial in-phase, rotation, and death) from the perspective of the interaction of toroidal vortices, particularly through the mechanism of vorticity reconnection. This work was focused on the simplest albeit sufficiently intriguing system consisting of three flames in an equal-lateral triangle arrangement. Consequently, we were able to depict the evolution of interacting toroidal vortices in great details. In order to avoid the numerical simulation of complex candle flames and to verify the existence of these dynamical modes in other flame systems, we adopted Bunsen-type burner of methane to produce flickering buoyant diffusion flames. We believed that the present work provide new insights to the study of multiple couple flames that are ubiquitous in nature, domestic applications, and industrial applications concerning flame stability and fire safety.

## 2. Computational Methods and Validations

In the present study, we were mainly concerned with the nonlinear dynamical behaviors of three identical flickering flames in an equal-lateral triangle arrangement. As illustrated in Fig. 1(a), three square-shaped Bunsen-type burners are of the same dimensions with a fixed edge length, $D = 10$ mm, and they are located at each vertex of the inscribed equal-lateral triangle of a circle of radius $R$. Methane gas is injected from the burner base with a uniform velocity, $U_0$, and goes through the adiabatic solid-wall burner with a height of $3D$. The computational domain of $16D \times 16D \times 24D$ has a uniform mesh of $160 \times 160 \times 240$, which has been validated by the domain and grid-independence studies [33]. The open boundary condition (i.e. gas flows into and out freely) was specified to all six surfaces of the computational domain. Following the previous work of Yang et al.[33], the mixing-limited chemical reaction model was adopted in the present simulations; radiation and soot formation were neglected for simplicity although we were fully aware of their quantitative effects[10, 39-46]. For all cases shown in the paper, the flames were in fully developed state as the simulation time was more than 20 times longer than the characteristic time, $2R/U_0$.

The open-source code, Fire Dynamics Simulator (FDS), was employed for simulating the unsteady, three-dimensional, incompressible (variable-density) flow with chemical heat release. The code has been widely used in fire dynamics problems in the past decade[47-53] and successfully applied to flickering buoyant diffusion flames [33]. More details on numerical methods and schemes refer to [54].

To validate the current numerical methods for well capturing the flickering phenomenon of buoyant jet diffusion flames, we conducted a series of simulations with variable $U_0$ and $g$ (the gravitational constant). It is seen in Fig. 1(b) that the present simulations well predict the well-known the scaling law for the flickering frequency, $f \sim \sqrt{g/D}$ [11, 55] and the previous experiments on jet flames[7, 56, 57]. Furthermore, the predicted flickering frequencies very slightly decrease with increasing $U_0$, confirming that the buoyancy is predominant for the diffusion flames concerned.

To facilitate the following discussion on vortex dynamics in the triple flame system, two example cases of single flickering flames are shown in Fig. 2. To visualize the toroidal vortex, we plotted the $Q$ criterion, where $Q = (|\Omega|^2 + |S|^2)/2$, $\Omega$ and $S$ are the anti-symmetric and symmetric components of the deformation rate tensor, $\nabla \vec{u}$.. The $Q$ criterion defines a vortex as a region where the magnitude of vorticity is greater than that of the strain rate. In the present study, the toroidal vortex is represented by the vorticity line that cross the highest Q-value.

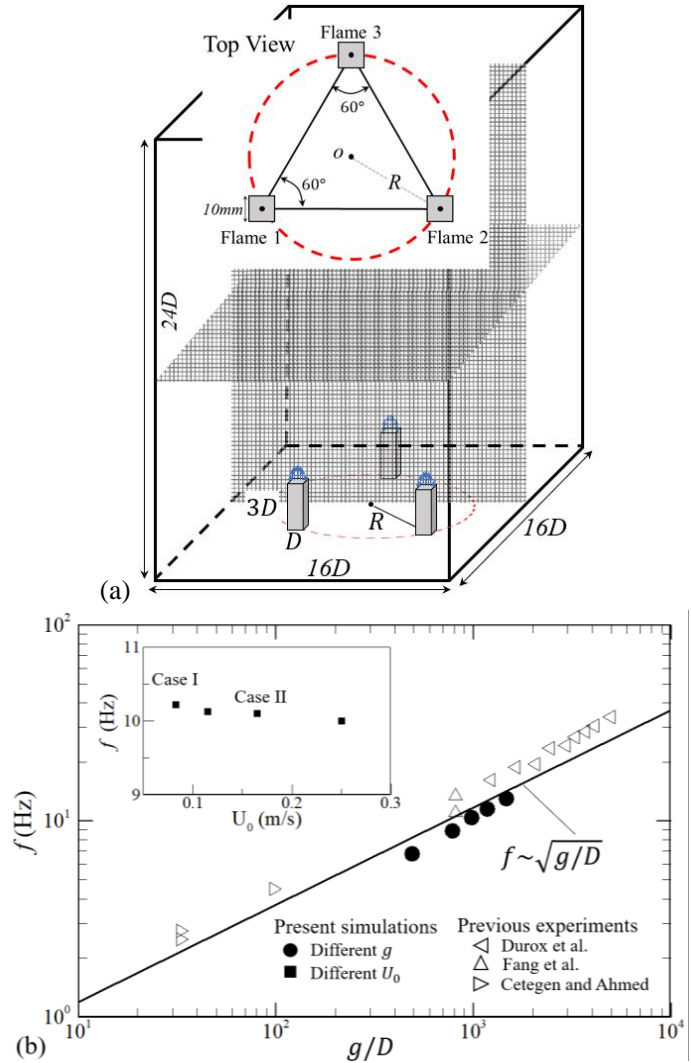

Fig. 1. (a) Schematic of the computational domain, mesh, and flame arrangement (top view); (b) Validation of computational methods for single flickering jet flames. Case I and Case II are also shown in Fig. 2.

The periodic behavior of a flickering flame and its associate toroidal vortices is clearly seen in Fig. 2. At the time instant $t_1$, a toroidal vortex is growing and rolling up, and the preceding toroidal vortex (already shed off) is moving downstream. The Q-contours well describe the vorticity distribution around the vortices. From $t_2$ to $t_3$, the toroidal vortex around the flame is contracted inwardly, and the flame is progressively necked. At the time instant $t_4$, the fully developed toroidal vortex is shed off, and a flame bubble is pinched off from its anchored flame. At the time instant $t_5$, a new toroidal vortex is generated at the flame base and will repeat its lifecycle. It is seen that the shed-off toroidal vortices can retain their shape and size but the vorticity fields around them are weakened due to viscous dissipation.

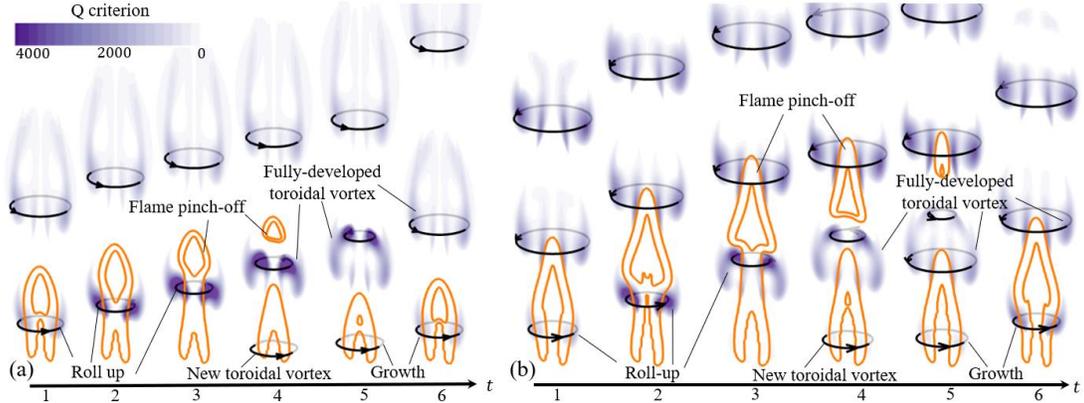

Fig. 2. The lifecycle of a toroidal vortex (denoted by the area of high positive Q criterion) generated by the flickering buoyant diffusion flame at (a) $U_0 = 0.165$ m/s, $Re = 50$, $Fr = 0.07$ and (b) $U_0 = 0.83$ m/s, $Re = 100$, $Fr = 0.28$, where $Re = U_0 D/\nu_F$ is the Reynolds number ($\nu_F = 1.65 \times 10^{-5}$ m$^2$/s is the kinematic viscosity of methane at room temperature) and $Fr = U_0^2/gD$ is the Froude number. The toroidal vortex is represented by the representative vortex line that crosses the highest Q-value. The flame shape is denoted by the orange contour line of heat release.

By comparing the two cases in Fig. 2, we noted that the decrease of $Fr$ significantly reduces the sizes of the flame and the flame bubble. Consequently, it is inferred that a sufficiently small $Fr$ can suppress the flame flicker [58, 59] or even cease the flame oscillation[6, 9]. Physically, this implies that the toroidal vortex requires a sufficiently large flame to grow to its critical circulation for shedding. This explains why many previous experiments used a bundle of small candles to create a bigger candle flame that can flicker.

## 3. Decoupled and Merging Modes of Triple Flickering Flames

When the flames are sufficiently away from each other, they flicker independently without coupling. The decoupled dynamical behavior is usually called as desynchronization. As a limiting case, the triple flame system at Re=100 and $R/D = 5$ are shown in Fig. 3. At such a large flame distance, the flames have very weekly interactions, and their dynamical mode is nearly decoupled. This can be seen in Fig. 3 that all toroidal vortices around flames keep their circular shape and the axial symmetry of each flame is hardly broken. In the further downstream, the shed-off toroidal vortices slightly deform due to the flow induced by the vortices to each other.

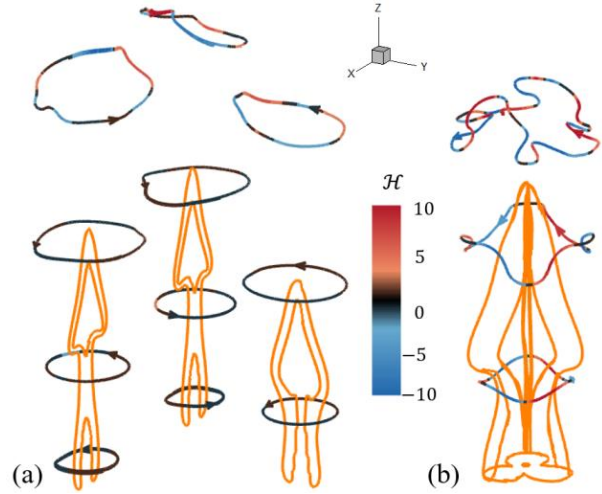

Fig. 3. The triple flame system at $Re = 100$ and (a) $R/D = 5$ and (b) $R/D = 1$. The representative vortex lines are colored by the normalized helicity density $\mathcal{H}$.

To quantify the vortex interaction through induced flow, we plotted the value of $\mathcal{H} = h/g$ along the vorticity lines, where $h = \vec{u} \cdot \vec{\omega}$ is the helicity density[60]. It should be noted that helicity (volumetric integral of $h$) is an important measure of the knottedness and/or linkage of the vortex lines in topological fluid dynamics because it is an invariant in an inviscid flow given no vorticity crossing at the boundary. The helicity density is equal to zero in an axisymmetric flow where $\vec{\omega}$ is orthogonal to $\vec{u}$. Consequently, a nonzero value of helicity density quantifies the non-orthogonality of $\vec{u}$ and $\vec{\omega}$, which is caused by the induced flow around a toroidal vortex by the other two. It is seen in Fig. 3 that the three circular vorticity lines retain on horizontal planes and have almost vanishing $\mathcal{H}$ before the three vortices are shed off. But the interaction of these shed-off vortices becomes stronger in the downstream as the vorticity lines are twisted and have non-zero $\mathcal{H}$.

As another limiting case, the triplet flame system at $Re = 100$ and $R/D = 1$ is shown in Fig. 3b. The three flames are completely merged to form a larger flickering flame with a smaller frequency because of the well-known scaling relation that the flickering frequency is inversely proportional of the square root of the flame burner size. This case is of little interest to the present study since there is no trace of separated toroidal vortices in the immediately downstream of burners.

## 4. In-phase and Flickering Death Modes of Triple Flickering Flames

At $Re = 100$ and $R/D = 1.6$, the triple flame system exhibits an in-phase synchronization, in which the three flames flicker with no phase difference, as shown in Fig. 4. The deformation, necking, and pinch-off of each flame qualitatively agree with the previous candle-flame experiment[34] (See Fig. S1 for more details). Regardless of some discrepancies compared with the experiment, the present simulation well captures the in-phase dynamical mode and verifies its existence in the triple jet diffusion flames.

To interpret this in-phase mode of the triple flame system, we can again refer to the evolution and interaction of toroidal vortices during one flickering period, as shown in Fig. 4. Specifically, the flame stretching, necking, and pinch-off are respectively associated with the growth, contraction, and shedding of a "trefoil" vortex during $t_3 \sim t_4$. The remaining question is how the "trefoil" vortex is formed. The evolution of three toroidal vortices during $t_1 \sim t_3$ suggests that the answer is the simultaneous connection of the vortices on the inner side of the flames.

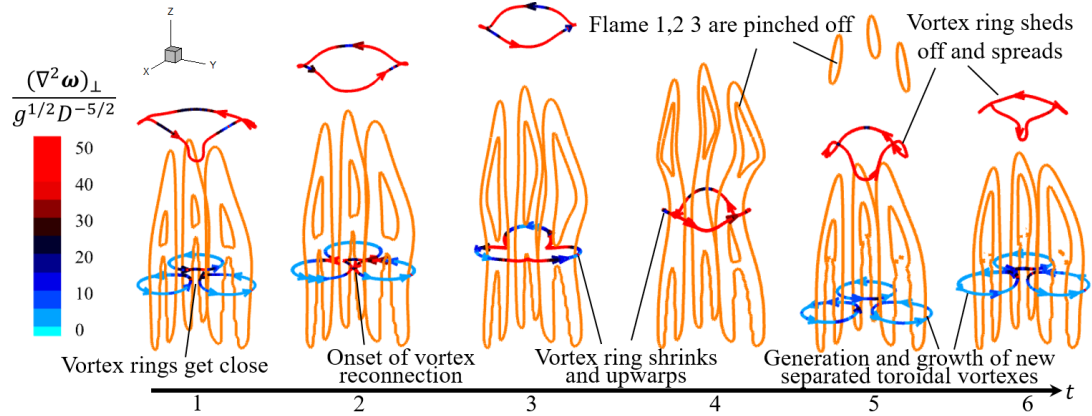

Fig. 4. The triple flame system at $Re = 100$ and $R/D = 1.6$. The representative vorticity lines cross the highest $Q$-value and are colored by the normalized $(\Delta^2 \omega)_\perp$.

Vortex interaction and particularly vortex reconnection is a long-standing problem in topological fluid dynamics[61, 62] owing to its important role in understanding both energy cascade and fine-scale mixing in turbulence [63-65]. According to Kida and Takaoka [insert reference], there are three distinct concepts of vortex reconnection: scalar, vortex, and vorticity reconnections, which are respectively referred to the change of topology of iso-surfaces of a passive scalar, of the vorticity magnitude, and of vorticity lines. Vortex reconnection in laboratory experiments is often visualized as a change of topology of a passive scalar like dye or smoke, which behaves quite differently from the vorticity field (either the vorticity magnitude or vorticity lines). In the present study, we adopted the concept of vorticity reconnection because it is valid only for viscous flows while the other two are possible in inviscid flows. In addition, we used the Q-criterion to identify vorticity lines in the present viscous flow, as discussed in Section 2.

Kida and Takaoka [insert reference] pointed out that helicity can be generated through vorticity reconnections but not vice versa, so helicity is not a good indicator for vorticity reconnection. According to the Helmholtz vortex theorem, the motion of vorticity lines is frozen in an inviscid flow and their reconnection is forbidden. Consequently, the breakdown of the frozen motion of vorticity lines are due to the viscous diffusion of vorticity, $\nu\nabla^2\vec{\omega}$. The natural decomposition of $\nu\nabla^2\vec{\omega}$ with respect to the direction of vorticity vector yields the parallel component $\nu(\nabla^2\vec{\omega})_\parallel$, which represents the stretching rate of vorticity lines, and the normal component, $\nu(\nabla^2\vec{\omega})_\perp$, which represents the deviation rate of vorticity lines and therefore a good indicator for vorticity reconnection.

Based on the above understanding, we calculated $(\nabla^2\vec{\omega})_\perp = (\Delta\vec{\omega} \times \vec{n}) \times \vec{n}$, where $\vec{n} = \vec{\omega}/|\vec{\omega}|$, normalized it by $\sqrt{g/D^5}$, and plotted it along each vorticity lines in Fig. 4. As the three toroidal vortices grow up, they get closer to each other, and the tendency of their reconnection becomes stronger as indicated as the red line segments for large values of normalized $(\nabla^2\vec{\omega})_\perp$ at the time instant $t_2$. Subsequently, the three vorticity lines cut out and reconnect to form a "trefoil" vorticity line, indicating the onset of vorticity reconnection at the time instant $t_3$. It is also noted that the "trefoil" vorticity line bends upwardly along the three flame directions to form a "saddle-trefoil" shape. This bending trend is clearly indicated by the large values of $(\nabla^2\vec{\omega})_\perp$.

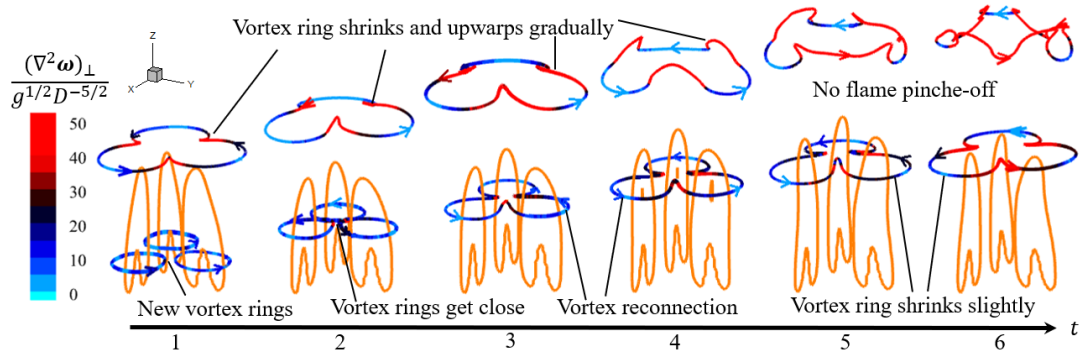

Fig. 5. The triple flame system at $Re = 50$ and $R/D = 1.6$. The representative vorticity lines cross the highest $Q$-value and are colored by the normalized $(\Delta^2\omega)_\perp$.

A flickering death mode (three flames cease to flicker but slightly oscillate) was observed in the triple flame system at the same $R/D = 1.6$ but a smaller $Re = 50$, as shown in Fig. 5. Compared with the in-phase mode shown in Fig. 4, the similar vorticity reconnection occurs to the three toroidal vortices, but the formed "trefoil" vortex does not have significant contracting and bending deformation before it sheds off from the flame (See Fig. S2 for more details). As the result, there is no significant flame necking and flame "bubble" pinch-off, rendering the cease of flame flicker. Because the flames still have a sight periodic oscillation with no phase difference, the flickering death mode can be treated as a special case of in-phase mode. It should be noted that the complete death mode (three flames cease to oscillate and fall into the steady combustion) is possible by completely suppressing the vortex shedding through further reducing $Re$ and therefore enhancing viscous dissipation.

5. **Rotation and Partial in-phase Modes of Triple Flickering Flames**

Compared with the in-phase and flickering death modes discussed in Section 4, the rotation and partial in-phase modes do not preserve the $D_3$ symmetry of the equal-lateral triangle flame system, and they show the features of spontaneous symmetry breaking.

At $Re = 100$ and $R/D = 2.0$, the triple jet diffusion flame system exhibits a rotation mode, in which the three flames alternatively flicker with a constant phase difference of $2\pi/3$. The deformation, necking, and pinch-off of each flame qualitatively agree with the previous candle-flame experiment[34] (See Fig. S3 for more details).

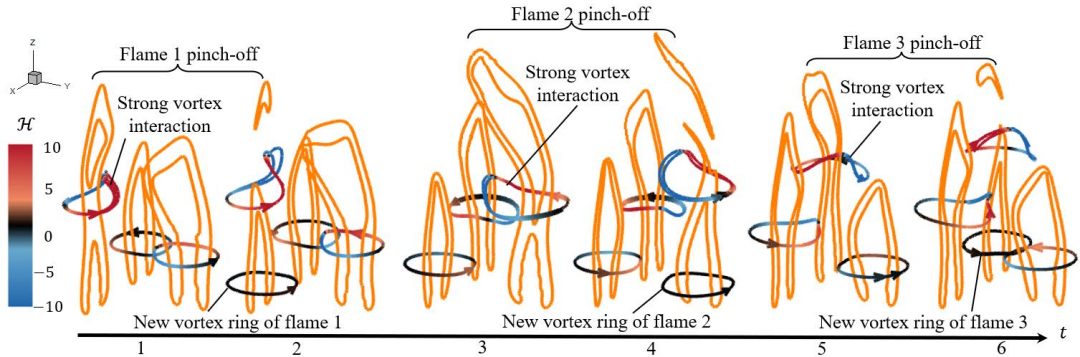

Fig. 6. The triple flame system at $Re = 100$ and $R/D = 2$. The vorticity lines cross the highest $Q$-value and are colored by the normalized helicity density $\mathcal{H}$.

The evolution and interaction of three toroidal vortices are shown in Fig. 6. It is clearly seen that there is no apparent vorticity reconnection between the three representative vorticity lines, which cross the local highest Q-value. At the time instant $t_1$, the height of the three toroidal vortices around the flames is in the descending order of 1–2–3 (corresponding to Flame 1, 2 and 3); Flame 1 has more significant stretching and necking compared with the other two. At the time instant $t_2$, the vortex of Flame 1 has shed off and a new vortex is formed at the flame base; the height of the three vortices is now in the descending order of 2–3–1. Subsequently, the order of the height of the vortices becomes 3–1–2 as the result of old vortex shedding and new vortex formation of Flame 2 (see the figures at time instants $t_3$ and $t_4$). The order of the height of the vortices returns to 1–2–3 as the result of old vortex shedding and new vortex formation of Flame 3 (see the figures at time instants $t_5$ and $t_6$). As such a complete period of the rotation mode is finished.

To quantify the vortex interaction in the rotation mode, we plotted the normalized helicity density $\mathcal{H}$ along each vorticity lines since there is no apparent vorticity reconnection and therefore the normalized $(\Delta^2 \omega)_\perp$ is not an appropriate quantity. It can seen that the change of the $\mathcal{H}$-value along each vorticity line follows the same order with that of the height of the vortex. For example, at the time instant $t_1$, the magnitude of $\mathcal{H}$-value along vortex 1 is the highest ($\mathcal{H}$ is mostly positive), that of vortex 2 is in the middle ($\mathcal{H}$ is mostly negative), and that of vortex 3 is the smallest ($\mathcal{H}$ is around zero). Physically, the vortex-induced flow is different around each vortex, and the positive (negative) $\mathcal{H}$ probably indicates that the induced flow enhances (suppresses) the vortex growth. Consequently, the shedding of old vortex 1 at the time instant $t_2$ relieves tits suppression on the following vortex 2, which in turn sheds off at $t_4$ to relieves its suppression on vortex 3.

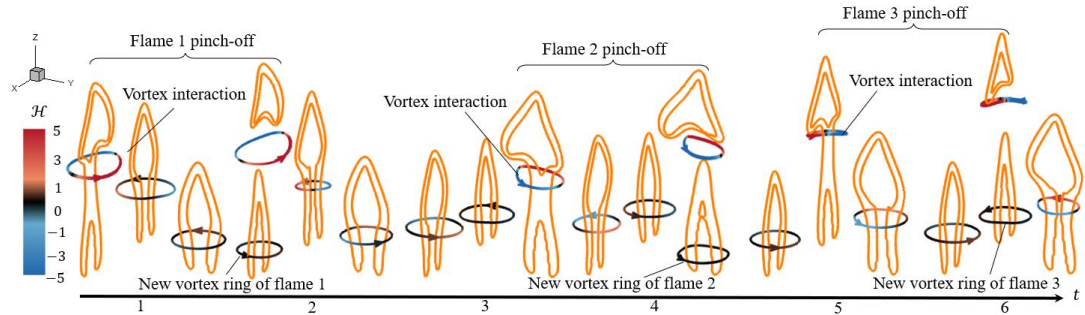

Fig. 7. The triple flame system at $Re = 100$ and $R/D = 3$. The representative vorticity lines cross the highest $Q$-value and are colored by the normalized helicity density $\mathcal{H}$.

The rotation mode can be observed at other flame distance, for example at $Re = 100$ and $R/D = 3$, as shown in Fig. 7. This case shows the similar phenomena of alternative flame flicker, and the major difference is the weaker vortex interaction due to the larger flame separation distance. This can be clearly seen from the normalized helicity density along each toroidal vortices.

Further increasing the flame distance to $R/D = 3.6$, we reproduced the partial in-phase mode (See Fig. S4 for more details)., in which two flames flicker without phase difference but another one flickers with a phase difference of $\pi$ (i.e. anti-phase). As shown in Fig. 8, vorticity reconnection occurs between two toroidal vortices so that their flame flicker in an in-phase synchronization. It should be noted that, due to the large $R/D$, the reconnection does not occur between the two representative vorticity lines that cross the highest Q-value. Instead, the vorticity lines that are defined by a smaller Q-value are used to denote the reconnection. The connected "8-shape" vortex contracts inwardly to cause the "necking" of the two flames to form two flame "bubbles", which pinch off from the anchoring flames due to the shedding of the "8-shape" vortex. This vortex shedding relieves the suppression of vortex-induced flow on the growth of the toroidal vortex around the third flame. This vortex subsequently grows and sheds off to cause the anti-phase flicker of the third flame.

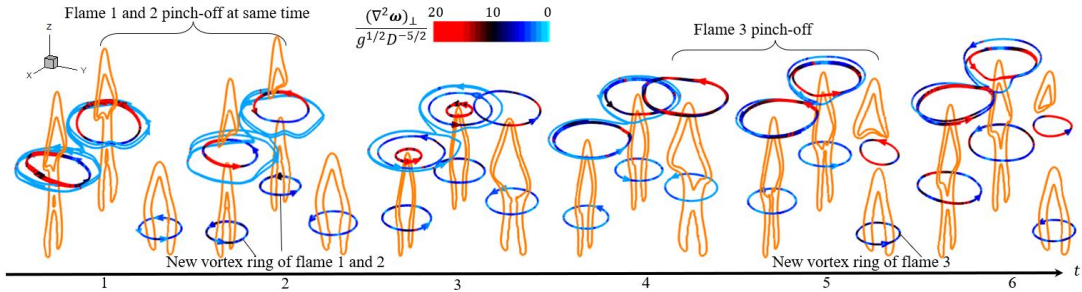

Fig.8. The triple flame system at $Re = 100$ and $R/D = 3.6$. The representative vorticity lines crosses the highest $Q$-value and are colored by the normalized $(\Delta^2 \omega)_\perp$.

## 6. Concluding Remarks

Multiple coupled flickering flames as a dynamical system of oscillators have gained increasing attentions in recent years. For the simplest albeit sufficiently interesting system consisting of three flickering diffusion flames in an equal-lateral triangle arrangement, two symmetric dynamical modes (i.e. the in-phase and death modes) and two asymmetric dynamical modes (i.e. the rotation and partial in-phase modes) was discovered in previous candle-flame experiments [insert reference], but the physical understanding toward these modes was far from being adequate.

In this study, we aimed to interpret the four distinct dynamical modes from the perspective of vortex interaction because the crucial role of buoyance-induced toroidal vortices has been substantiated in many previous experimental and theoretical studies. We computationally reproduced the four dynamical modes for the triple flickering buoyant jet diffusion flames of methane in the equal-lateral triangle arrangement at different $R/D$. The vortex-dynamical interpretation to the modes is summarized as follows.

In the in-phase mode occurring at relatively small $R/D$, the reconnection of three toroidal results in a larger trefoil vortex, whose periodic shedding causes the in-phase flicker of the three flames. The occurrence of the death mode requires smaller flames and hence smaller $Re$ because the trefoil vortex may not be able to reach its critical circulation for vortex shedding if the viscous dissipation is sufficiently strong. In the rotation mode occurring as relatively larger $R/D$, the vorticity reconnection of three presentative vorticity lines that across the highest Q-value does not happen. Instead, each toroidal vortex alternatively reaches the highest vertical height and sheds off from its anchoring flame. The interaction between the toroidal vortices is through the vortex-induced flow, whose influence can be quantified by the nonzero helicity density around the vortices. The shedding the first vortex relieves the suppression of the vortex-induced flow on the second vortex, which in turns grows up, sheds off, and relieves its suppression on the third vortex. In the partial in-phase mode, a vorticity reconnection occurs between two toroidal vortices and hence their flames flicker without phase difference. The shedding of the "8-shape" vortex relieves the suppression of the vortex-induced flow on the third vortex, which subsequently grows up and sheds off to result in the anti-phase flicker of its anchoring flame.

It is noted that the present study answered the questions how the triplet flickering flame system exhibits four distinct dynamical modes and how the interaction between toroidal vortices interpret these modes. However, we fully recognized that the present study barely addresses the questions why the triplet flame system in the $D_3$ symmetry has these modes and how these modes transition between each other. Future works addressing these remaining equations will gain us a deeper understanding of the physical origin and transition of the dynamical modes in the system of multiple flickering flames.

**Acknowledgements**

This work is financially supported by the National Natural Science Foundation of China (No. 52176134) and by the Hong Kong Polytechnic University (G-UAHP).

**Supplementary material**

Supplementary material is available at.